\documentclass{aastex}
\usepackage{emulateapj5}

\begin{document}

\def\'#1{\ifx#1i{\accent"13\i}\else{\accent"13#1}\fi}

\def\BP{Ballesteros-Paredes}
\def\Pext{P_{\rm ext}}
\def\Pint{P_{\rm int}}
\def\rhom{\bar\rho}
\def\VS{V\'azquez-Semadeni}

\slugcomment{Submitted to The Astrophysical Journal}

\shorttitle{Can Hydrostatic Cores Form Within Isothermal Molecular Clouds?}
\shortauthors{V\'azquez-Semadeni, Shadmehri \& Ballesteros-Paredes}

\title{Can Hydrostatic Cores Form Within Isothermal Molecular Clouds?}

\author{Enrique V\'azquez-Semadeni$^1$, Mohsen Shadmehri$^2$ and Javier
\BP$^1$} 
\affil{$^1$Instituto de Astronomia, UNAM, Apdo. Postal 72-3 (Xangari),
Morelia, Michocana 58089, M\'exico
\email{e.vazquez, j.ballesteros@astrosmo.unam.mx}} 
\affil{$^2$Department of Physics, School of Science, Ferdowsi
University, Mashhad, Iran} \email{mshadmehri@science1.um.ac.ir}

\begin{abstract}
Under the assumptions that molecular clouds are nearly spatially and
temporally isothermal and that the density peaks (``cores'') within them
are formed by turbulent fluctuations, we argue that cores cannot reach a
hydrostatic (or magneto-static) state as a consequence of their formation
process. In the non-magnetic case, this is a consequence of the fact
that, for cores at the same temperature of the clouds, the necessary
Bonnor-Ebert truncation at a finite radius is not feasible, unless it
amounts to a shock, which is clearly a dynamical feature, or the core is 
really embedded in hotter gas. Otherwise,
quiescent cores must have non-discontinuous density profiles until they
merge with their parent cloud, constituting extended structures. For these,
%
%
we argue that any equilibrium configuration with non-vanishing
central density is unstable. Since the cores' environment (the
molecular cloud) is turbulent, no reason exists for them to settle
into an unstable equilibrium. Instead, in this case, cores must be
dynamical entities that can either be pushed into collapse, or else 
``rebound'' towards the mean pressure and density as the parent
cloud. Nevertheless, rebounding cores are
delayed in their re-expansion by their own self-gravity. We give a crude
estimate for the re-expansion time as a function of the closeness of the final
compression state to the threshold of instability, finding typical
values $\sim 1$ Myr, i.e., of the order of a few free-fall times. Our
results support the notion that not all 
cores observed in molecular clouds need to be on route to forming stars, but
that instead a class of ``failed cores'' should exist, which must eventually
re-expand and disperse, and which can be identified with observed starless
cores. In the magnetic case, recent observational and theoretical work
suggests that all cores are critical or supercritical, and are thus
qualitatively equivalent to the non-magnetic case. This is, however, not 
a problem for the efficiency of star formation: within the turbulent scenario
the low efficiency of star formation does not need to rely on magnetic
support of the cores, but instead is a consequence of the low
probability of forming collapsing cores in a medium that is globally
supported by turbulence. Our results support the notion that the entire
star formation process is dynamical, with no intermediate hydrostatic stages.

\end{abstract}

\keywords{ISM: structure - stars: formation - hydrodynamics -
turbulence} 

\section{Introduction}\label{sec:intro}

One of the most important goals in the study of star formation is to
understand the state and physical conditions of the molecular cloud
cores from which the stars form. The prevailing view concerning
low-mass-star-forming cores is that they are quasi-static equilibrium
configurations supported against gravitational collapse by a
combination of magnetic, thermal and turbulent pressures (e.g.,
Mouschovias 1976a,b; Shu, Adams \& Lizano 1987). When considering only thermal
pressure, two variants of the equilibrium structures are usually
discussed: either singular isothermal structures, with diverging
central densities and smooth $r^{-2}$ density dependence extending to
infinity (e.g., Shu et al.\ 1987), or finite-central density
structures, truncated at some finite radius and confined by the
pressure of some external medium, generally assumed to be at higher
temperatures and lower densities than the isothermal core (Ebert 1955;
Bonnor 1956). More recently, the equilibria of non-axisymmetric
configurations have also been studied (e.g., Fiege \& Pudritz 2000;
Curry 2000; Galli et al. 2001; Shadmehri \& Ghanbari 2001; Lombardi \&
Bertin 2001; Curry \& Stahler 2001).

The support from magnetic fields is generally included through the
consideration of the mass-to-magnetic flux ratio of the core, since,
assuming that the latter has a fixed mass, the flux freezing condition
implies that its mass-to-flux ratio is constant (Chandrasekhar \& Fermi
1953; Mestel \& Spitzer 1956). Under isothermal conditions, the magnetic 
pressure and the gravitational energy scale as the same power of the
core's volume; thus, self-gravity cannot overcome the magnetic support
if the mass-to-flux ratio is smaller than some critical value, and
collapse can only occur as the magnetic flux diffuses out of the cloud
by ambipolar diffusion (see, e.g., Mestel \& Spitzer 1956; Mouschovias \&
Spitzer 1976; Shu, Adams \& Lizano 1987).

On the other hand, it is well established that the molecular clouds
within which the cores form are turbulent, with linewidths that are
supersonic for scales $\gtrsim 0.1$ pc (e.g., Larson 1981), and with
(magnetohydrodynamic) turbulent motions providing most of the support
against gravity, with only a minor role of thermal pressure at all but
the smallest ($\lesssim 0.1$ pc) scales. 
Thus, there appears to be a conceptual gap between the turbulent nature
of the clouds and the quasi-hydrostatic assumed nature of the cores.
The cores in molecular clouds must be subject to
global motions and distortions, as well as mass exchange with its
surroundings (in general, to continuous ``morphing''), and, in fact,
are likely to be themselves the turbulent density fluctuations within
the clouds (von Weizs\"acker 1951; Bania \& Lyon 1980; Scalo 1987;
Elmegreen 1993; \BP, \VS\ \& Scalo 1999, hereafter BVS99; Padoan et al.\
2001). At present, one interpretation is that the cores are the
dissipative end of the turbulent cascade, because the velocity
dispersion within them becomes sonic or subsonic (e.g., Goodman et al.\
1998). However, in actuality, substructure is seen down to the smallest
resolved scales (e.g., Falgarone, Puget \& P\'erault
1992), and appears even within what were previously considered to be
``smooth'' cores, as the resolution is improved (Wilner et al.\ 2000). Also,
inflow motions, themselves with substructure, are generally seen around
these cores (e.g. Myers, Evans \& Ohashi 2000).
Moreover, if the transonic cores are part of a compressible cascade,
they do not need to be the dissipative end of it, but may simply mark
the transition to a regime of nearly incompressible turbulence (\VS, \BP
\& Klessen 2002, 2003). 

This issue also poses a problem for the idea of confining
clumps by turbulent pressure, since the latter is in general
anisotropic and transient at large scales. In this regard,
it is worth remarking that a frequent interpretation of the
role of turbulent pressure in ``confining'' cores is that the total
thermal-plus-turbulent pressure is larger outside a core than inside
it, because the turbulent velocity dispersion increases with size. This
is, however, an incorrect interpretation, as the dependence of turbulent 
pressure with size scale is a non-local property referring to statistical
averages over domains of a given size, not to a gradient of the local
value of the velocity dispersion as larger distances from the
core's center are considered.

If the density peaks (clumps and cores) within molecular clouds have
a dynamic origin, then an immediate question is whether they can
ever reach hydrostatic equilibrium. Several pieces of evidence suggest
that this is not possible. First, Tohline et al.\ (1987) considered
the potential energy curve of an initially gravitationally-stable
fluid parcel in a radiative medium characterized by an effective
adiabatic (or ``polytropic'') exponent, showing that it has a
``thermal energy barrier'' that must be overcome, say by an increase
in the external turbulent ram pressure, in order to push the parcel
into gravitational collapse. In particular, these authors estimated
the Mach numbers required for this to occur. Although those authors
did not discuss it, the production of a hydrostatic configuration
within this framework would require hitting precisely the tip of such
``barrier'', the probability of which is vanishingly small, because
the tips of potential barriers constitute unstable equilibria. 

Second, although Shu
(1977) has argued that the singular isothermal sphere is the state
asymptotically approached by the flow as it seeks to establish detailed
mechanical balance when its parts can communicate subsonically with one
another, the maintenance of this configuration for long times seems
highly unlikely, as this configuration constitutes an {\it unstable}
equilibrium, being the precursor of gravitational collapse. If the
formation of the core is a dynamical process, no reason exists for the
flow to relax onto an unstable equilibrium. Such a state can be used 
as an initial condition in simulations of gravitational collapse, but
does not represent itself a realistic state that can be reached by a gas 
parcel in a turbulent medium.

Third, Clarke \& Pringle (1997) have pointed out 
that cores cool mainly through optically thick lines, but are heated by
cosmic rays, and therefore may be dynamically unstable, as velocity
gradients may enhance local cooling. 

Fourth,
numerical simulations of self-gravitating, turbulent clouds (e.g., \VS\
et al.\ 1996; Klessen, Heitsch \& Mac Low 2000; Heitsch, Mac Low \&
Klessen 2001; Bate et al.\ 2002) never show the production of hydrostatic
objects. Instead, once a fluid parcel is compressed strongly enough to
become gravitationally bound, it proceeds to collapse right away.
Specifically, BVS99 suggested that hydrostatic structures cannot be formed by
turbulent compressions in polytropic flows, in which the pressure is
given by $P \propto \rho^\gamma$, where $\rho$ is the mass density and
$\gamma$ is the effective polytropic exponent. This is because the
collapse of an initially{\it  stable} gas parcel can only be induced
(i.e, the parcel made unstable) by a
(strong enough) mechanical compression if $\gamma < \gamma_{\rm c}$,
where the value of $\gamma_{\rm c}$ depends on the dimensionality of
the compression and the specific heat ratio of the gas (see, e.g.,
\VS, Passot \& Pouquet 1996). However, once collapse has been
initiated, it cannot be halted unless $\gamma$ changes in the process,
to become larger than $\gamma_{\rm c}$ again. In other words, for
non-isothermal situations, with $\gamma > \gamma_{\rm c}$, equilibria 
can be found even if the external pressure is time variable.  This is
why stars can be formed as stable entities 
from highly anisotropic, dynamic, time-dependent accretion
(Hartmann, \BP\ \&  Bergin 2001).  For systems that are much closer
to isothermal, such as molecular cloud cores, the boundary pressures
are indispensible in establishing stable equilibria, which are therefore 
not expected to exist in an isothermal turbulent medium with a
fluctuating ram pressure. In fact, an analysis of the energy contents of
the clouds in numerical simulations 
shows that they are in near energy equipartition but nowhere near virial
{\it equilibirum} (VE) (see Ballesteros-Paredes \& V\'azquez-Semadeni 1995;
1997; Shadmheri, V\'azquez-Semadeni \& Ballesteros-Paredes 2003). This
suggests that observations of rough energy equipartition (e.g., Myers
\& Goodman 1988) does not necessarily imply that clouds are in such detailed
mechanical balance.

The only case when numerical studies show the formation of
(magneto)static structures occurs in simulations of super-Jeans, yet
subcritical clouds (e.g., Ostriker, Gammie \& Stone 1999), in which the
whole box is subcritical. However, as we discuss in \S \ref{sec:magn},
we believe that this 
is an artifact of the simulations being performed in closed boxes that
do not allow further mass accretion until the system becomes supercritical.

In this paper, we provide further arguments against the possibility of
molecular cloud cores being hydrostatic entities, and argue in favor of
them being instead transients, although with low (subsonic) internal velocity 
dispersion. The plan of the paper is as follows: In \S
\ref{sec:trunc_ext} we argue against the possibility of truncated
Bonnor-Ebert-type configurations arising in nearly single-temperature
molecular clouds, suggesting instead that cores must either be
shock-confined or else have smooth(extended) density profiles, and then
discuss the stability of extended structures, noting that unstable
equilibria are not expected to arise in turbulent media. In \S
\ref{sec:re-exp} we give a crude estimate of the re-expansion time of
density peaks (cores) that are not sufficiently compressed to undergo
gravitational collapse. In \S \ref{sec:magn} we then discuss the
magnetic case, arguing that the subcritical case is also just a
transient, on the basis of previous results existing in the
literature. Then, in \S \ref{sec:dyn_scen} we discuss how the proposed
dynamical nature of the cores is not inconsistent with observations, and 
finally, in \S \ref{sec:conclusions}, we summarize our results
and give some conclusions.

\section{The non-magnetic case} \label{sec:non_magn}

\subsection{Truncated versus extended configurations}
\label{sec:trunc_ext} 

The notion that molecular cloud cores are 
nearly-hydrostatic structures can probably be traced back to the
classical work of Ebert (1955) and Bonnor (1956). These authors
independently studied the
stability {\it against gravitational collapse} of truncated isothermal
configurations (Bonnor-Ebert, or BE, spheres) bounded by a wall or by
a tenuous hot medium capable of maintaining pressure balance at the
sphere's boundary but without contributing appreciably to the
self-gravity of the system. 
The BE analysis shows that such
structures are stable for $\xi \lesssim 6.5$, where $\xi \equiv r/\left[2
\pi L_{\rm J}(\rho_{\rm c})\right]$ is a nondimensional radial variable,
normalized by the Jeans length at the central density $\rho_{\rm
c}$. A less remembered fact is 
that the presence of the wall or hot confining medium is indispensible
in order to prevent instability towards {\it reexpansion}. Indeed, a
simple application of the Virial Theorem to the case of a hydrostatic
self-gravitating sphere in the {\it absence} of a confining medium
shows that this case is always unstable for media obeying a polytropic
equation ($P \propto \rho^\gamma$) with $\gamma <4/3$ (which includes
the isothermal case, $\gamma=1$). Upon a perturbation 
in its volume, such a hydrostatic sphere will 
engage in either runaway expansion or runaway contraction (collapse). 

Now, since molecular clouds are quite close to being isothermal (e.g.,
Myers 1978; Pratap et al.\ 1997; Scalo et al.\ 1998),
the cores within them are essentially at the same temperature as their
``confining'' medium. Thus, no thermal discontinuity can occur analogous to
that required by the BE configuration. The only possibility for a
discontinuity within a single-temperature medium is for there to exist a
shock at the boundary, across which both the density and the pressure change 
discontinuously, with the pressure jump being supplemented by ram
pressure. In what follows, we do not discuss this possibility any further,
as it already amounts to a dynamic, rather than hydrostatic, situation.

We should note that another possibility exists for the realization of a
BE-type configuration, namely that the core is really ``trapped'' within
a hotter 
region, as is the case of the well known globule Barnard 68 (Alves, Lada 
\& Lada 2001). In this case, the surrounding medium can indeed confine
the globule without 
contributing significantly to the gravitational potential. However, in
the general molecular-cloud case of a confining medium at the same
temperature as the core, this possibility is excluded. Thus, we conclude 
that the cores within molecular clouds must either be shock-confined
(and thus transient), or have smooth density profiles, extending in
principle to infinity, asymptotically approaching a zero background
density. We refer to these as ``extended'' profiles. In practice,
molecular cloud cores have densities at least 100 times larger than the
average molecular cloud density, so the assumtion of vanishing
background density appears reasonable.

We now wish to consider the stability of extended density
configurations. The best-known of these is the singular isothermal
sphere, which has an $r^{-2}$ density profile, and is known to be
gravitationally unstable (see, e.g., Shu 1977). Since we are interested
in density 
fluctuations produced by turbulent compressions, we require them to
have a finite central density. These configurations can be obtained by
numerically integrating the Lane-Emden equation to arbitrarily large
radii. However, this implies that they are equivalent to a BE sphere of
arbitrarily large radius, and are thus gravitationally unstable (Bonnor
1956)\footnote{Again, in practice, a
central-to-background density ratio of $\sim 100$ is well into the
unstable regime for BE-spheres.} (we thank D.\ Galli for suggesting this
argument). We conclude 
that {\it all} extended hydrostatic configurations in a
single-temperature medium are gravitationally unstable.

It is worth comparing this situation with that of the BE configurations, 
which have a well-defined range of ratios of central-to-boundary
densities for which the configuration is stable. This occurs because the
confinement of the sphere by a hot 
tenuous medium circumvents the need to satisfy hydrostatic
equilibrium at all distances from the center; i.e., the hydrostatic
condition is only imposed out to the ``boundary'' radius. We see that
the existence of a stable range of BE spheres is precisely allowed by
the truncation. 

In the standard picture of low-mass star formation, some
equilibrium state of the kind described above is
usually taken as the initial condition for subsequent collapse. However, 
we see that, since extended configurations are unstable, no reason exists
for them to ever be reached if they are originated by dynamic turbulent
compressions.\footnote{Note that Hunter (1977) has shown that states
unstable under the BE criterion need not readily collapse, but may 
undergo large-amplitude radial oscillations. However, his initial
conditions were very different than those expected from turbulent
formation of the cores (our basic working hypothesis here): while he was 
(again) taking the hydrostatic state as the initial condition and then
perturbed it, in a turbulent cloud core formation is expected to occur
dynamically, so that the static situation is never realized. Only if the 
hydrostatic state is a {\it stable} equilibrium can it then become an
{\it attractor} of the dynamical evolution, and cause the static configuration 
to appear.} Moreover, as we have discussed, truncated configurations
are inapplicable within single-temperature media. Thus, the above arguments
are suggestive that 
{\it the initial conditions for star formation should be dynamical in
general, rather than quasi-static.}

\subsection{Re-expansion time of ``failed'' compressions} \label{sec:re-exp}

The arguments above would seem to suggest that all density fluctuations
within a turbulent molecular cloud should collapse. However, it is easy
to see that this need not be the case, if one is willing to relax the
requirement of hydrostatic equilibrium. For example, consider any stable 
BE sphere, i.e., of radius smaller than the critical one, and surround
it with a medium with a steeper density profile than that of
equilibrium, say a power law $r^n$, with $n< -2$, but maintaining
pressure continuity. In this case, the
configuration has less than one Jeans mass at every radius, and will
proceed to re-expansion. Thus, if such a configuration is produced by a
turbulent compression, its fate is to re-expand, after the compression
subsides. 

It is important to note, however, that the re-expansion process must
occur on a longer time scale than the compression because of the retarding
action of self-gravity. This implies that cores that are compressed to
conditions close to those of instability are expected to have somewhat
longer lifetimes than those that proceed to collapse, and thus suggests
that possibly the majority of the cores observed in molecular clouds 
are {\it not} on their way to collapse, but rather towards re-expansion
and merging into their ambient medium. It is thus of interest to
estimate their extended lifetimes. 

\begin{figure}
\plotone{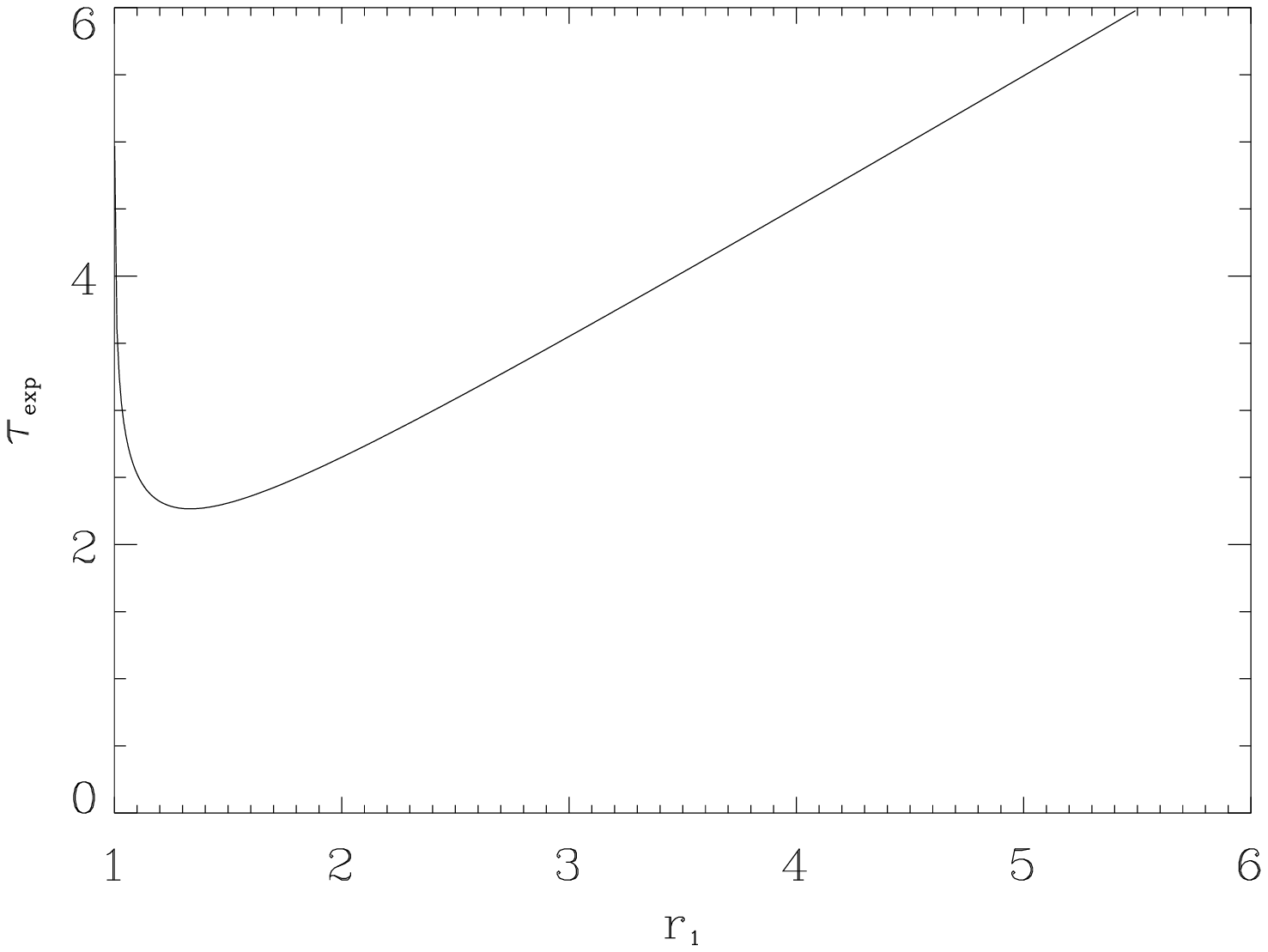}
\caption{Re-expansion time of a core, in units of the free-fall time,
defined as the time necessary to 
double the initial core's radius, as a function of the initial radius
$r_1$, normalized to the equilibrium radius.} 
\label{fig:re-exp}
\end{figure}

A crude estimate of the re-expansion time can be given in terms of the
Virial Theorem (VT), because in this case we are interested in the
characteristic growth time of an unstable equilibrium configuration. The 
VT allows the description of this situation, in particular of the case in
which the evolution is towards re-expansion, by consideration of a
gas sphere in equilibrium between its self-gravity and internal pressure 
exclusively. As mentioned above, this configuration is unstable, and can 
evolve either towards collapse or re-expansion upon a perturbation of
the sphere's volume. Thus, the VT in this case provides a reasonable
approximation to the case of an unstable extended configuration. We can
then proceed as follows. The VT for an isothermal spherical gas mass
(``cloud'') of volume $V$ and mean density $\rhom$ in empty space is
\begin{equation}
\frac{1}{2} \ddot I= 3 Mc^2 - \alpha GM^2/R,
\label{eq:VT}
\end{equation}
where the overhead dots indicate time derivatives, $M=\int_V \rho dV$ is
the cloud's mass, $R$ is its radius, 
$I=\int_V \rho r^2 dV$ is its moment of inertia, $c$ is the
sound speed, and $\alpha$ is a factor of order unity. We can obtain an
evolution equation for the cloud's radius by 
replacing the radius-dependent density by its mean value in the
expression for $I$, to find $I\approx MR^2$. Thus, 
\begin{equation}
\frac{1}{2} \ddot I \approx M\left[R(t) \ddot R(t) + \dot
R^2(t)\right].
\label{eq:I_R}
\end{equation}
Equating equations (\ref{eq:VT}) and (\ref{eq:I_R}), we obtain
\begin{equation}
\left[R(t) \ddot R(t) + \dot R^2(t)\right] = 3c^2 -\alpha GM/R.
\label{eq:R_evol}
\end{equation}
This equation can be integrated analytically, with solution
\begin{eqnarray}
\tau = & \frac{1} {\sqrt 3}\biggl[\sqrt{(r_2-1)^2-(r_1-1)^2} + \nonumber \\
 & \ln \left(\frac{r_2 -1+\sqrt{(r_2-1)^2 - (r_1-1)^2}}{r_1-1}\right) \biggr]
\label{eq:sol_R_evol}
\end{eqnarray}
where $r_1$ and $r_2$ are the initial and final radii of expansion,
normalized to the
equilibrium radius $R_{\rm e}=\alpha GM/3 c^2$, and 
$\tau=t/t_{\rm ff}$ is the time, non-dimensionalized to the
free-fall time $t_{\rm ff}=R_{\rm e}/c$. The characteristic
re-expansion time can be defined as the time required to double the
initial radius (i.e., $r_2 = 2 r_1$), starting from an initial condition
$r_1>1$. Figure \ref{fig:re-exp} shows this characteristic time as a
function of $r_1$. We see that when $r_1$ is very close to unity
(i.e., linear perturbations from the equilibrium radius), the
re-expansion time can be up to a few times the free-fall
time. Moderately nonlinear perturbations have the shortest
re-expansion times, because the initial force imbalance is greater,
yet the final size is still not much larger than twice the equilibrium
radius. Finally, for larger initial radii (far from the equilibrium
value), the re-expansion time approaches that of free expansion at the
sound speed. We conclude that the re-expansion time is at least larger
than twice the free-fall time, making the probability of observing a
core in this process larger by this factor than that of observing a
free-falling core, in agreement with the fact that molecular clouds
are generally observed to contain more 
starless than star-forming cores (e.g., Taylor, Morata \& Williams
1996; Lee \& Myers 1999; see also Evans 1999 and references therein).

\section{The magnetic case} \label{sec:magn}

In the magnetic case, the classical Virial-Theorem (VT) analysis
(Chandrasekhar \& Fermi 1953; 
Spitzer 1968; Mouschovias 1976a,b; Mouschovias \& Spitzer 1976; Zweibel
1990) predicts the existence of sub- and super-critical 
configurations (Shu et al.\ 1987; Lizano \& Shu 1989) depending on
whether the mass-to-magnetic flux ratio is below or above a critical
value $(M/\phi)_{\rm c}$. Subcritical configurations are known not to
be able to collapse gravitationally unless the magnetic flux is lost
by some process such as ambipolar diffusion. Supercritical
configurations, on the other hand, are analogous to the non-magnetic
case, except for the fact that the cloud behaves as if having an
``effective'' mass, reduced by an amount equal to the critical mass
(which depends on the magnetic field strength). 
The VT analysis, however, has the same problem as the BE
treatment in that it neglects to satisfy the hydrostatic condition
beyond the cloud radius, and so it is not applicable for cores within
clouds at their same temperature. We are thus faced again with the need to
consider extended configurations.

%
%
The equilibria of magnetically supported cores is significantly more
complicated than that of non-magnetic ones because of the anisotropy
introduced by the field, and the many possible field geometries. 
Instability analyses of extended magneto-static layers and cylinders with
a variety of 
geometrical field configurations have been performed by many workers
(e.g., Chandrasekhar \& Fermi 1953; Nakamura, Hanawa \& Nakano 1991, 1993; 
Nakajima \& Hanawa 1996; Gehman, Adams \& Watkins 1996; Nagai, Inutsuka
\& Miyama 1998). The layers are in general unstable, although long-lived 
intermediate filamentary states in route to collapse have been reported
(Nakajima \& Hanawa 1996). Nevertheless, it is clear that in a turbulent 
medium, there is no reason for the unstable equilibrium configurations
to arise. The structures of greatest interest here may be the
intermediate, long-lived structures mentioned above, arising not from
the collapse of magneto-static initial states, but of
dynamically-produced structures.

Moreover, additional considerations suggest that the very concept of
subcritical cores may not be realized in practice within molecular
clouds if the latter are supercritical as a whole, as it appears to be both
from obervations (Crutcher 1999; Bourke et al.\ 2001; Crutcher, Heiles
\& Troland 2002) and from theoretical considerations (Nakano
1998). Indeed, the VT treatment 
giving stability below the critical mass-flux ratio assumes a 
fixed mass for the cloud or core under consideration. Instead, a
core that forms part of a larger cloud, has a mass that is not fixed,
and continued accretion along magnetic field lines can occur until the
core becomes supercritical (Hartmann et al.\ 2001). 

This possibility appears to be supported by recent numerical simulations
of MHD turbulent flows (Padoan \& Nordlund 1999; Ostriker, Stone \&
Gammie 2001; Passot \& \VS\ 2002), which have shown that the magnetic field 
is essentially decorrelated from the density. Passot \& \VS\ (2002) have
explained this phenomenon in terms of the different scalings of the
magnetic pressure with density for the different MHD wave modes, and
shown that for the slow mode the magnetic pressure can actually be
{\it anti-}correlated with the density. Additionally, numerical
simulations in which the entire computational domain is supercritical
systematically show the 
collapse of the {\it local} density peaks (Heitsch, Mac Low \&
Klessen 2001), while magnetostatic cores are not observed (R.\ Klessen,
2002, private communication). Only when the entire computational box is
artificially constrained to be subcritical by the (usually periodic)
boundary conditions (which do not allow further accretion along field
lines) the collapse of both the large and the small scales is
prevented (e.g., Ostriker et al.\ 1999), giving rise to flattened
structures, and having led some groups to consider
two-dimensional models of molecular clouds (e.g., Shu et al.\
1999). However, if accretion were allowed from the surrounding medium, a 
supercritical configuration can eventually be reached, provided the entire
cloud is supercritical, in order for there to be enough mass available.
These considerations suggest that the
subcritical state is in fact a transient stage prior to the formation
of supercritical structures that can subsequently collapse.

\section{The dynamic scenario} \label{sec:dyn_scen} 

The suggestion that molecular cloud cores cannot be in hydrostatic
equilibrium immediately raises two questions: One, how should we then
interpret the low (subsonic) velocity dipersions observed within the
cores? Two, if the time scale for external pressure variations were
sufficiently large, should we not then expect quasi-hydrostatic
cores that are hydrostatic for all practical purposes?

The answer to the first question can be found in the scenario outlined
in BVS99 and \VS\ et al. (2002, 2003). This is based on the suggestion that
turbulence plays a dual role in structures from giant molecular clouds
to cloud clumps (Sasao 1973; Falgarone et al.\ 1992; \VS\ \&
Passot 1999; Klessen et al.\ 2000), in such a way that it contributes to
the global support
while promoting fragmentation into smaller-scale structures. The process
is hierarchical, repeating itself towards smaller scales (Kornreich \&
Scalo 2000) until the turbulent velocity dispersion within the next
level of structures becomes subsonic, as dictated by a turbulent cascade
in which smaller have smaller velocity dispersions. At this
point, no further sub-fragmentation can occur, because subsonic
isothermal turbulence cannot produce large density fluctuations, and
morever it becomes sub-dominant in the support of the structure (Padoan
1995). In this scenario, a core is made by an initially supersonic
velocity fluctuation {\it at larger scale}, but during the process the
compression slows down, because of generation of smaller-scale motions,
dissipation, and transfer to internal energy, which is however readily
radiated away. Thus, in this scenario, subsonic cores (some of them
collapsing, some re-expanding) are a natural outcome and the ending
point of the compressible, lossy turbulent cascade. Work is in progress
for the formulation of a quantitative model.

Concerning the second question, it is important to remark that, in order 
to make a significant density fluctuation from a turbulent compression,
the latter must have an appreciable ($>1$) {\it external} Mach
number.\footnote{The ``external'' qualification means that this is the Mach
number of the turbulent compression that makes the core. Inside the 
core, the velocity dispersion is expected to be reduced by the
dissipation that ensues from the shocks produced. This is what is meant
by a ``lossy'' compressible cascade.} Therefore, the formation of the
cores is expected to occur on short time scales, of the order of the
crossing time of the next large scales in the hierarchy (Elmegreen
1993), and the possibility of long time scales for the pressure
variations is excluded.

\section{Conclusions} \label{sec:conclusions}

In this paper we have argued that the final state of isothermal fluid parcels
compressed into ``cores'' by turbulent velocity fluctuations cannot
remain in equilibrium. In the non-magnetic case, this is due to the
isothermality of the flow, 
which implies that the a continuous pressure profile requires a continuous
density profile, except if it is supplemented by ram 
pressure in a shock. Since shocks are already non-hydrostatic features,
thus agreeing with our claim, we focus on continuous-profile
(``extended'') structures. For these, we argued that all equilibrium
configurations are unstable, contrary to the stability range found for
truncated BE-type structures, and thus are not expected to arise in a
dynamic, fluctuating medium. Thus, cores must in general collapse or
re-expand, but cannot remain in equilibrium, unless they happen to enter 
(or be ``captured'' in) a hotter region, as is the case of the
much-discussed B68 globule. In the magnetic case, we have recalled
several recent results  
suggesting that all cores are critical or supercritical, thus being
qualitatively equivalent to the non-magnetic case regarding their
possibility of collapse.

Although our arguments are conceptually very simple, we believe they have
been overlooked in the literature because the hydrostatic state is
normally considered as an {\it initial} condition, accepted without
questioning how such state can be arrived at, and because the turbulent 
pressure is implicitly assumed to be ``microscopic'' (i.e., of
characteristic scales much smaller than the core), neglecting the fact
that molecular clouds are globally turbulent and that the bulk of the
turbulent energy is at the largest scales, as clearly suggested by
the observed velocity dispersion-size scaling relation (Larson 1981),
implying that the cores themselves {\it are} the turbulent density
fluctuations. 

%

Our results have the implication that many observed cores are not on
route to forming stars, but instead ``fail'', and must re-expand and
merge back into the general molecular cloud medium. For these,
the re-expansion time is expected to be larger than the compression
time due to the retarding action of self-gravity. A simple estimate
based on virial balance suggests that the re-expansion time is of the
order of a few free-fall times. This is
consistent with the facts that molecular clouds typically contain more
starless than star-forming cores (e.g., Taylor, Morata \& Williams
1996; Lee \& Myers 1999; see also Evans 1999 and references therein),
and that most of the cores do not appear to be gravitationally bound
(e.g., Blitz \& Williams 1999). 

It is worthwhile to note that these time scales are over one order of
magnitude shorter than estimates based on ambipolar diffusion (see,
e.g., McKee et al.\ 1993).  Indeed, the long ambipolar diffusion time
scales were necessary to explain the low efficiency of star formation in
the old hydrostatic paradigm, but in the dynamic scenario of star
formation, the low efficiency is a natural consequence that only a small
fraction of the mass in a molecular cloud is deposited by the turbulence
in collapsing cores (Padoan 1995; \VS\ et al.\ 2002, 2003), and does not
need to rely on magnetic support of the cores.


We thus suggest that hydrostatic
configurations have no room in the process of star formation in
turbulent, isothermal molecular clouds. Theories of core structure and
star formation should consider the fact that core formation is a
dynamical process. This probably implies that the density profile in
cores is a function of time, and therefore {\it not unique}. This may be in
agreement with the fact that recent surveys find {\it distributions} of
the scaling exponent, rather than clearly defined unique values (e.g.,
Shirley et al.\ 2002). Another implication is that fundamental
properties like the star formation efficiency may be {\it statistical}
consequences of the turbulence in molecular clouds (Elmegreen 1993;
Padoan 1995; V\'azquez-Semadeni et al.\ 2002), rather than depending on
ambipolar diffusion to break the equilibrium state.

\acknowledgements

We have benefitted from comments and criticisms by Shantanu Basu,
Daniele Galli, Susana Lizano and Lee Mundy. We especially thank Lee
Hartmann, for valuable suggestions for the contents and presentation.
We also thank the anonymous referee for an exceptionally deep report
(including plots and calculations!)
which showed holes in the arguments presented in the initial version of
the paper, prompting us to find a more direct argumentation.
We acknowledge partial financial support from CONACYT grants
27752-E to E.V.-S and I 39318-E to J. B.-P., and from Ferdowsi
University to M.S. This work has
made extensive use of NASA's Astrophysics Abstract Data Service.

\clearpage



\end{document}